\RequirePackage{rotating}
\documentclass[nonlinenumbers]{aastex631}

\usepackage{newtxtext,newtxmath}
\usepackage{url}
\usepackage[T1]{fontenc}

\DeclareRobustCommand{\VAN}[3]{#2}
\let\VANthebibliography\thebibliography
\def\thebibliography{\DeclareRobustCommand{\VAN}[3]{##3}\VANthebibliography}

\usepackage{graphicx}	
\usepackage{amsmath}	

\usepackage[graphicx]{realboxes}
\usepackage{float}

\begin{document}

\title{Active Asteroid 311P/PANSTARRS: Rotational Instability as the Origin of Multi-tails?}

\author{Bin Liu}
\affiliation{School of Aeronautics and Astronautics, Shenzhen Campus of Sun Yat-sen University, Shenzhen, Guangdong 518107, China }
\affiliation{Shenzhen Key Laboratory of Intelligent Microsatellite Constellation, Shenzhen Campus of Sun Yat-sen University, Shenzhen, Guangdong 518107, China }

\author{Xiaodong Liu}
\affiliation{School of Aeronautics and Astronautics, Shenzhen Campus of Sun Yat-sen University, Shenzhen, Guangdong 518107, China }
\affiliation{Shenzhen Key Laboratory of Intelligent Microsatellite Constellation, Shenzhen Campus of Sun Yat-sen University, Shenzhen, Guangdong 518107, China }

\author{Xiaoyu Jia}
\affiliation{Beijing Institute of Spacecraft System Engineering, Beijing 100094, China }

\author{Fei Li}
\affiliation{Beijing Institute of Spacecraft System Engineering, Beijing 100094, China }

\author{Yuhui Zhao}
\affiliation{Purple Mountain Observatory, Chinese Academy of Sciences. Nanjing 210023, China }

\author{LiangLiang Yu}
\affiliation{State Key Laboratory of Lunar and Planetary Sciences, Macau University of Science and Technology, Macau, China }

\correspondingauthor{Xiaodong Liu}
\email{liuxd36@mail.sysu.edu.cn}

\begin{abstract}

The active asteroid 311P is one of the two targets of a planned Chinese asteroid exploration mission Tianwen-2. During 2013, 311P experienced several mass-loss events and exhibited multiple comet-like tails.
Here we analyze the morphology and surface brightness of the tails to investigate the dust environment around the nucleus and mechanism of activities. We enhance the features of the tails using image processing techniques to obtain information about the morphology of the tails, and fit processed images to the syndyne-synchrone diagrams. The fitting results give estimations of the upper limits of the durations ($2\sim8$ days) of eruptions and the dust size range ($0.006\sim38.9$ mm) in the tails. The results of surface photometry performed for each dust tail show that the brightness distribution index of each tail ranged from approximately -1.81 to 0 and the dust size distribution indices of 311P’s tails ranged from -2.29 to -1.45. The quantity of particles in each tail ranged from 0.5 to 8 $\times10^6$\,kg, which leads to a total dust-loss quantity of $3.0\times10^7$\,kg and a mass loss rate of 1.59 kg s$^{-1}$. Sublimation, continuous impacts or tidal forces of planets are excluded as explanations for 311P's activities, and rotational instability remains a possible activation cause without strong evidence against it.
\end{abstract}

\section{Introduction} \label{sec:intro}

It was conventionally considered that comets and asteroids are different types of small solar system bodies \citep{Jewitt2022}. 
In recent years, however, the discoveries of active asteroids have blurred the boundary line between asteroids and comets.
Active asteroids usually exhibit detectable mass loss in confined segments of their orbits. More than 40 known active asteroids have been discovered at the time of this writing. There are several different physical mechanisms to explain such mass-loss activity, including water sublimation (e.g.~1 Ceres \citep{kuppers2014localized} and 288P \citep{agarwal2017binary}), impact ejecta (e.g.~354P/LINEAR (\citealt{kleyna2013p}; \citealt{kim2017anisotropic}) and P/2016 G1 \citep{moreno2016early}),
rotational instability (e.g.~331P/Gibbs \citep{drahus2015fast} and P/2013 R3 \citep{jewitt2014disintegrating}),
and thermal effects (e.g.~3200 Phaethon \citep{jewitt2013dust}). 
Activated asteroids display comet-like tails which consist of a large number of dust grains emitted from the nucleus. The information about mass-loss mechanisms and the physical properties of dust particles could be inferred by analyzing dust tails of active asteroids.

In this paper, we focus on the active asteroid 311P (also known as P/2013 P5), which is an interesting target for space exploration. The orbital parameters of 311P (collected from the JPL Small-Body Database) are as follows: semi-major axis $a=2.189$\,AU, eccentricity $e=0.116$, and inclination $i=4.968\degr$, yielding a Jupiter Tisserand parameter $T_\mathrm{J}=3.66$. 311P was discovered on 18 August 2013 by the Pan-STARRS1 telescope \citep{bolin2013comet}. In follow-up observations by using the Hubble Space Telescope, it was found that 311P generated multiple comet-like tails,
suggesting multiple outbursts occurred in early 2013 (\citealt{jewitt2013extraordinary}; 
\citealt{jewitt2015episodic}).
Rotational disruption \citep{jewitt2015episodic} or rubbing of binary components \citep{hainaut2014continued} is the most likely driven mechanism for 311P's activities. The observations by \citet{jewitt2018nucleus} strongly suggest that 311P is a binary system, which implies that the activities of 311P may be caused by rubbing of the binary components.

The absolute magnitude of the active asteroid 311P measured by \citet{jewitt2018nucleus} with the Hubble Space Telescope is $H_\mathrm{v}=19.14\pm0.02$, and the photometry of its nucleus implies that the radius of 311P is $r_\mathrm{e}=(0.19\pm0.03)$ km. 
The lower limit of the rotation period of 311P is about 5.4 hours, which is derived from the light curve of 311P observed by the Hubble Space Telescope \citep{jewitt2018nucleus}. \citet{hainaut2014continued} estimated the peak times of the activities and total mass production around the activity peaks by analyzing images from the Canada-France-Hawaii Telescope, the Perkins Telescope, the New Technology Telescope, and the TRAnsiting Planets and PlanetesImals Small Telescope-South. \citet{moreno2014intermittent} analyzed the observation images obtained by the Hubble Space Telescope as well as the Gran Telescopio CANARIAS, modeled a three-month activity resulting from spin instability, and estimated the total dust mass, which is of the order of $10^{7}$ kg. Furthermore, \citet{moreno2014intermittent} indicated that isotropic ejection model does not fit 311P's activities based on the Hubble Space Telescope data. After estimating the initial time of each ejection, \citet{jewitt2015episodic} measured the total dust quantity around the nucleus, and obtained the upper and lower limits of grain sizes and the surface brightness profiles of the tails by analysing data from the Hubble Space Telescope. Moreover, the activity mechanism of 311P was also discussed in \citet{jewitt2015episodic}.

In this paper, we perform data reduction and reanalysis on the observation data from the HST program 13609 (PI: David Jewitt), which was previously analyzed by \citet{jewitt2015episodic}. Our work provides new aspects on activities and tails of 311P, which are shown as follows. The length of each tail at each observation epoch is obtained (Table \ref{tab:2}). The upper limit of the duration of each emission activity (Table \ref{tab:tab3}) is estimated by analyzing the widths of the tails. The number of the dust particles for different grain radii (Figure \ref{fig:fig6}) and the dust mass integrated along the radial distance of the tail from the nucleus (Figure \ref{fig:fig5}) in each tail are determined by applying the Finson-Probstein theory to the surface photometry result. For most of the tails of 311P, we observed that (Figure \ref{fig:fig9}) the dust size distributions are steeper for the tails with larger average grain size, and their indices are close to that of the power-law distribution of the self-organized critical sandpiles. We also discuss the effect of the tidal forces of planets (Figure \ref{fig:fig7}b), which is ruled out as the origin of activity.

We also revisit some aspects that were previously addressed by \citet{jewitt2015episodic}. 
We use the same method of fitting the synchrones to tails' position angles to derive the starting times of ejections (Table \ref{tab:tab3}), the results of which differ a bit from \citet{jewitt2015episodic} because we use the boundaries of the tails for fitting instead of the brightest lines. We estimate the minimum grain size in each tail by fitting the tails' boundaries with the syndynes, the method of which is different from \citet{jewitt2015episodic} where the minimum grain size was determined from the tail's length but leads to consistent results. We adopt the same method to obtain the tail's surface brightness profile (Figure \ref{fig:fig4}) for each tail in all observed epochs, and our results are generally consistent with \citet{jewitt2015episodic} where the surface brightness profiles of certain selected tails were shown. Besides, we discuss and rule out
the possibilities of sublimation (Figure \ref{fig:fig7}a) and continuous impacts as activity mechanisms, and show that rotational instability remains a possible activation cause of 311P, which is consistent with \citet{jewitt2015episodic}.

The forthcoming Chinese deep space mission Tianwen-2 (previously known as ZhengHe mission) will explore 311P \citep{zhang2021china}, which is planned to be launched around 2025. Onboard the Tianwen-2 spacecraft there will be a dust analyzer, which is capable of physical analysis of dust particles \citep{zhao2022design}. Research on the dust environment of 311P is helpful and important for science and mission planning of Tianwen-2.

The paper is organized as follows. Archival images and data reduction are described in Section \ref{sec:Observations}.
An analysis of the dust tails morphology by using the syndyne-synchrone diagram is presented in Section \ref{sec:DustTailMorphology}.
We analyze the surface brightness of the dust tails in Section \ref{sec:TailSurfaceBrightness}.
In Section \ref{sec:QuantityOfDust}, we estimate the total amount and the size distribution of dust contained in each tail. The possible activity mechanisms of 311P are discussed in Section \ref{sec:Discussion}. Finally, the conclusions of this paper are summarized in Section \ref{sec:Conclusions}.
    \begin{table*}
	    \caption{Observation geometries of 311P/PanSTARRS}
	    \label{tab:tab1}
	    \begin{tabular}{lccccccccc}
		\hline
	    Date & DOY$^a$ & R$^b$ & $\Delta^{c}$ & $\alpha^d$ & PsAng$^e$ & PsAMV$^f$ & $\nu^{g}$ & $\delta^{h}$ & Telescope\\
		\hline
		30 August 2013 & 242 & 2.13 & 1.13 & 5.5 & 202.1 & 245.2 & 277.4 & -3.7 & CFHT\\
		1 September 2013 & 244 & 2.12 & 1.12 & 4.8 & 191.8 & 245.1 & 278.1 & -3.8 & NTT\\
		3 September 2013 & 246 & 2.12 & 1.12 & 4.3 & 178.3 & 245.1 & 278.1 & -3.9 & NTT\\
        5 September 2013 & 248 & 2.12 & 1.12 & 4.1 & 162.4 & 245.0 & 279.3 & -4.0 & CFHT\\
        6 September 2013 & 249 & 2.12 & 1.12 & 4.0 & 153.0 & 245.0 & 279.7 & -4.0 & NTT\\
        10 September 2013 & 253 & 2.11 & 1.11 & 4.8 & 125.0 & 244.9 & 281.0 & -4.2 & CFHT\\
        10 September 2013 & 253 & 2.11 & 1.11 & 4.8 & 125.0 & 244.9 & 281.0 & -4.2 & HUBBLE\\
        11 September 2013 & 254 & 2.11 & 1.11 & 5.2 & 119.5 & 244.8 & 281.3 & -4.2 & CA1.23m\\
        23 September 2013 & 266 & 2.10 & 1.14 & 10.7 & 89.2 & 244.5 & 285.2 & -4.3 & HUBBLE\\
        25 September 2013 & 268 & 2.09 & 1.14 & 11.7 & 87.1 & 244.4 & 285.9 & -4.3 & CA1.23m\\
        28 September 2013 & 271 & 2.09 & 1.15 & 13.2 & 84.5 & 244.4 & 286.9 & -4.3 & CFHT\\
        29 September 2013 & 272 & 2.09 & 1.16 & 13.7 & 83.8 & 244.3 & 287.2 & -4.3 & CFHT\\
        30 September 2013 & 273 & 2.09 & 1.16 & 14.2 & 83.1 & 244.3 & 287.5 & -4.3 & CFHT\\
        4 October 2013 & 277 & 2.08 & 1.18 & 16.0 & 80.7 & 244.2 & 288.9 & -4.2 & PERKINS\\
        5 October 2013 & 278 & 2.08 & 1.18 & 16.4 & 80.2 & 244.2 & 289.2 & -4.2 & TRAPPIST\\
        7 October 2013 & 280 & 2.08 & 1.20 & 17.3 & 79.3 & 244.2 & 289.9 & -4.1 & GTC\\
        18 October 2013 & 291 & 2.06 & 1.27 & 21.5 & 75.6 & 244.1 & 293.6 & -3.8 & HUBBLE\\
        8 November 2013 & 312 & 2.04 & 1.45 & 26.8 & 71.2 & 244.1 & 300.8 & -2.7 & GTC\\
        13 November 2013 & 317 & 2.03 & 1.50 & 27.6 & 70.4 & 244.2 & 302.5 & -2.5 & HUBBLE\\
        8 December 2013 & 342 & 2.01 & 1.75 & 29.4 & 67.3 & 244.6 & 311.4 & -1.1 & HUBBLE\\
        31 December 2013 & 365 & 1.98 & 2.00 & 28.6 & 65.4 & 245.4 & 319.8 & 0 & HUBBLE\\
        11 February 2014 & 407 & 1.95 & 2.38 & 23.9 & 64.5 & 248.8 & 335.4 & 1.4 & HUBBLE\\        
        13 November 2014 & 686 & 2.11 & 2.29 & 25.6 & 291.5 & 295.7 & 78.2 & -1.5 & HUBBLE\\
        3 March 2015 & 792 & 2.25 & 1.30 & 9.9 & 313.9 & 293.3 & 110.6 & 3.3 & HUBBLE\\
        19 March 2015 & 808 & 2.27 & 1.28 & 3.8 & 15.7 & 294.1 & 115.1 & 3.8 & HUBBLE\\
        7 April 2015 & 827 & 2.29 & 1.34 & 10.2 & 93.8 & 294.9 & 120.4 & 3.7 & HUBBLE\\
        4 May 2015 & 854 & 2.32 & 1.57 & 20.0 & 107.4 & 295.4 & 127.8 & 2.6 & HUBBLE\\
        29 June 2015 & 910 & 2.38 & 2.26 & 25.2 & 114.1 & 294.3 & 142.5 & 0.1 & HUBBLE\\
        27 July 2015 & 938 & 2.40 & 2.60 & 23.0 & 115.1 & 293.0 & 149.6 & -0.8 & HUBBLE\\
		\hline
		\multicolumn{10}{l}{$^a$ Day of year, 2013/01/01 = 1.}\\
		\multicolumn{10}{l}{$^b$ Heliocentric distance, in AU.}\\
		\multicolumn{10}{l}{$^c$ Geocentric distance, in AU.}\\
		\multicolumn{10}{l}{$^d$ Phase angle, Sun-311P-Earth, in degrees.}\\
		\multicolumn{10}{l}{$^e$ Position angle of the projected Sun-311P radial direction, in degrees.}\\
		\multicolumn{10}{l}{$^f$ Position angle of the projected negative 311P’s heliocentric velocity vector, in degrees.}\\
		\multicolumn{10}{l}{$^g$ True anomaly, in degrees.}\\
		\multicolumn{10}{l}{$^h$ Orbital plane angle (the angle between the observer and the orbital plane of 311P), in degrees.}\\
	\end{tabular}
    \end{table*}

\section{Observations Data}
\label{sec:Observations}

The observation data of 311P shown in this section cover a two-year period, starting from August 2013 to July 2015. We review the previous observation data of 311P by seven ground and space telescopes, including: the Canada-France-Hawaii Telescope, the TRAnsiting Planets and PlanetesImals Small Telescope-South, the New Technology Telescope, the Calar Alto Telescope, the Gran Telescopio CANARIAS, the Hubble Space Telescope, and Perkins. Table \ref{tab:tab1} summarizes the observing geometries of these observation data. It should be noted that only the data from the Hubble Space Telescope (Section \ref{section_Hubble}) are used for the analysis in this manuscript.

\subsection{Canada-France-Hawaii Telescope}
Images of 311P were obtained through the 3.6 m Canada-France-Hawaii Telescope at the top of Maunakea, Hawaii, with the MegaCam wide-field imager on 30 August, 5, 10, 28, 29 and 30 September of 2013 \citep{hainaut2014continued}.
The MegaCam wide-field imager consists of 40 CCDs with $2048\times4612$ pixel, with a resolution of $0.187\arcsec$ per pixel \citep{boulade2003megacam}.
CFHT observations of 311P were identified on six days in August and September 2013, all through a Sloan $g'$, $r'$, $i'$ filter. 
Exposure times ranged from 540 seconds to 1800 seconds.

The phase angle of 311P ranged from $4.1\degr$ to $14.2\degr$ during multiple observation epochs, while the heliocentric and geocentric distances ranged from $2.13$ to $2.09$\,AU, and $1.13$ to $1.16$\,AU, respectively. 
As shown in Figure 2a-c-e-h and Figure 3c of \citet{hainaut2014continued}, the composite images of the active asteroid 311P displayed a compact coma and three narrow and bright dust tails. 
However, the surface brightness of the tails steeply decreased over time, and the position angle of tails evolved rapidly within one month.

\subsection{The Perkins Telescope}
Observations were taken on UT 04 October 2013 at Lowell Observatory, by the Perkins 1.8 m diameters telescope, located in Arizona. 
The $2048\times2048$ charge-coupled device camera, PRISM, has a pixel scale of $0.39\arcsec$\,pixel$^{-1}$ \citep{janes2004first}. 
311P was observed for one night through a Bessel R band filter at phase angle of $16.4\degr$, 
at a heliocentric distance of 2.08 AU and a geocentric distance of 1.08 AU. A composite image of 311P, obtained by shifting images of exposure time 3600 s, is shown in Figure 2i of \citet{hainaut2014continued}. 
The tails of 311P became more dispersible on 04 October 2013 than those on 30 September 2013, and the tails in the images gradually weakened over time.

\subsection{TRAnsiting Planets and PlanetesImals Small Telescope-South}
On 5 October 2013, 311P was observed with the TRAPPISTCAM FLI ProLine PL-3041-BB of the 0.6 meter telescope at the La Silla European Southern Observatory, Chile. 
The receiver is the TRAPPISTCAM with a matrix of $2048\times2048$ pixels, a field-of-view of $22\degr\times22\degr$, a pixel scale of $0.64\arcsec$ per pixel, and a pixel size of 15 µm \citep{jehin2011trappist}.
The active asteroid 311P was observed via the Cousin R filter. 
The images of 311P were taken with a total exposure time of 20880 s.

\subsection{New Technology Telescope}
On 1, 3, and 6 September of 2013, observations of P/2013 P5 were made at the European Southern Observatory, Chile, with the $2048\times2048$ ESO$\#40$ CCD detector of the 3.56-m New Technology Telescope.
Binning $2\times2$ was used, resulting in an pixel scale of $0.24\arcsec$\,pixel$^{-1}$ and a field-of-view of $4.1\degr$ \citep{dekker1986eso}. 
Images were taken in the Bessel V and R bands, and exposure times of 3-nights observations are 4500 s, 13500 s and 10200 s, respectively.

\begin{figure}
\centering
    \includegraphics[width=0.7\columnwidth]{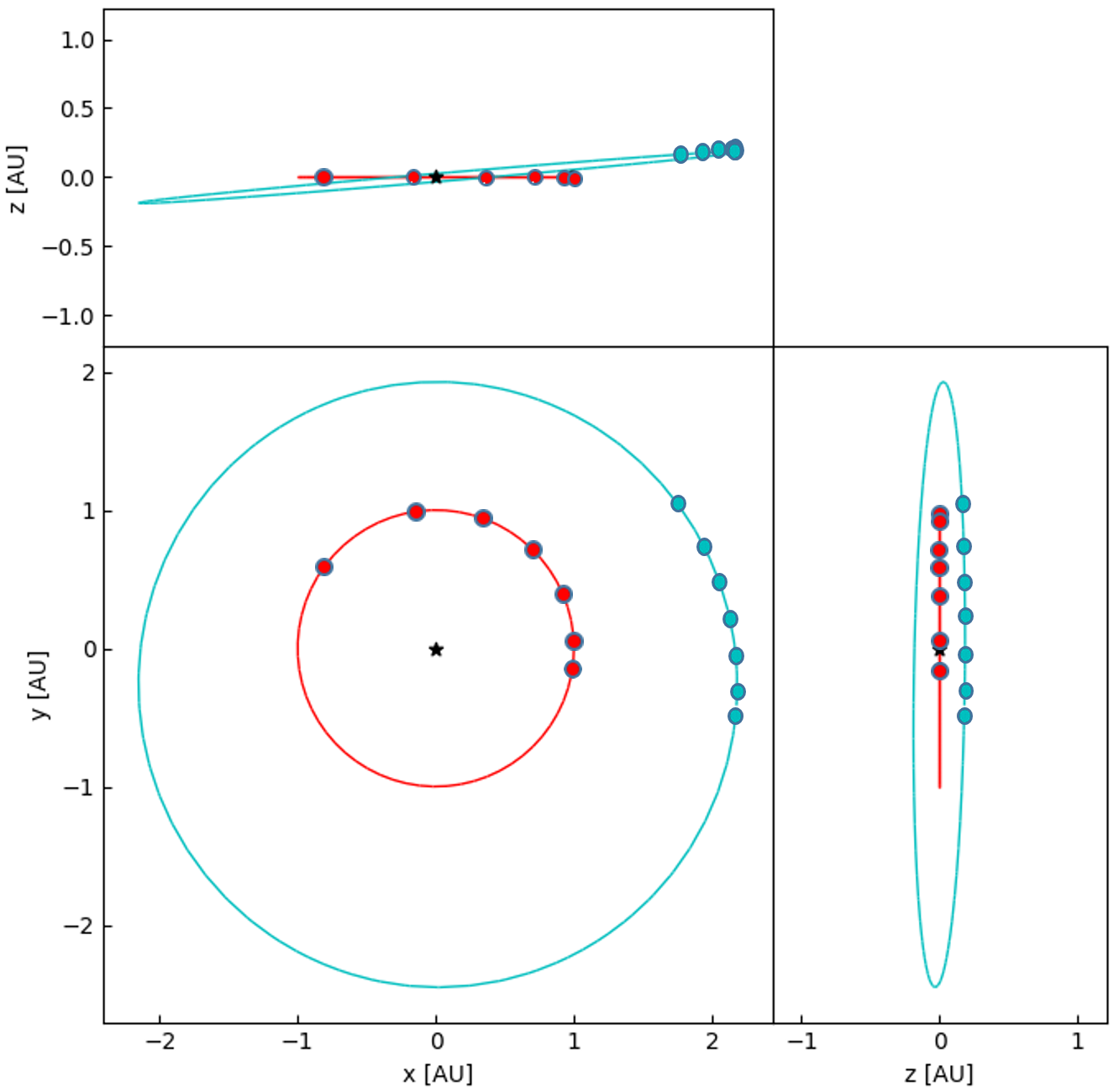}
    \caption{Projection of the orbit of 311P on the $xy$ plane, $xz$ plane, and $yz$ plane of the ecliptic coordinate system ECLIPJ2000. The outer cyan line corresponds to the orbit of 311P, and the inner one corresponds to the orbit of the Earth. The symbol $\star$ in the center represents the Sun. Cyan and red dots on different orbits represent the positions of 311P and the Earth at different observational epochs (10 September 2013, 23 September 2013, 18 October 2013, 13 November 2013, 8 December 2013, 31 December 2013, and 11 February 2014), respectively. The orbital simulation of 311P is performed with an N-body integrator REBOUND \citep{rein2011rebound}.}
    \label{fig:fig1}
\end{figure}

\subsection{Calar Alto Telescope}
On 05 October 2013, 311P was traced by \citet{hainaut2014continued} using the 1.23 m diameter Calar Alto Telescope, located on Calar Alto,
Almería province in Spain, and the DLR-KMIII camera. 
DLR-KMIII camera is equipped with $4$K$ \times4$K pixels e2v CCD231-84-NIMO-BI-DD charge-coupled device, giving an image scale of $0.628\arcsec$\,pixel$^{-1}$, 
and a field view of approximately $21.4\degr$ \citep{leinert1995measurements}. Johnson\_R filter was used twice with exposure times of 3300 s and 2100 s, respectively.

\subsection{Gran Telescopio CANARIAS}
To observe 311P, \citet{moreno2014intermittent} used the 10.4 m diameter Gran Telescopio CANARIAS (GTC), located on the island of La Palma, and Optical System for Image and Low Resolution Integrated Spectroscopy camera (OSIRIS). 
OSIRIS is equipped with $2048\times4096$ pixels charge-coupled device, 
giving an image scale of $0.127\arcsec$\, pixel$^{-1}$, and a field of view of about $7.8\arcmin\times7.8\arcmin$. Sloan $r$ and $g$ filters were used \citep{larkin2006osiris}. 
Two composite images from different dates are shown in Figure 3 of \citet{moreno2014intermittent}.

\subsection{The Hubble Space Telescope}
\label{section_Hubble}
The Hubble Space Telescope was used to observe 311P with Target-of-Opportunity time on 10 and 23 September 2013 (program number 13475, PI: David Jewitt; \citealt{jewitt2013extraordinary}), 18 October 2013, 13 November 2013, 8 December 2013, 31 December 2013, and 11 February 2014 (program number 13609, PI: David Jewitt; \citealt{jewitt2015episodic}). 
During all observing epochs, a total exposure time of 1973 s was obtained with the 2K subarray of the WFC3 camera with the pixel scale of $0.04\arcsec$\,pixel$^{-1}$ \citep{bouwens2010discovery}. 
Figure~\ref{fig:fig1} shows the position of the Earth-Sun-311P in the ecliptic coordinate system ECLIPJ2000. 
311P passed the perihelion at a perihelion distance of 1.93 AU on 16 April 2014, and all of Hubble observations were launched before perihelion. 

The publicly accessible observations of 311P are acquired from the Mikulski Archive for Space Telescopes (MAST) (see details of MAST in \citet{padovani1998multi}). Data reduction and frame adding are completed with the Astroart software \citep{nicolini2003astroart}.

\begin{figure*}
    \plotone{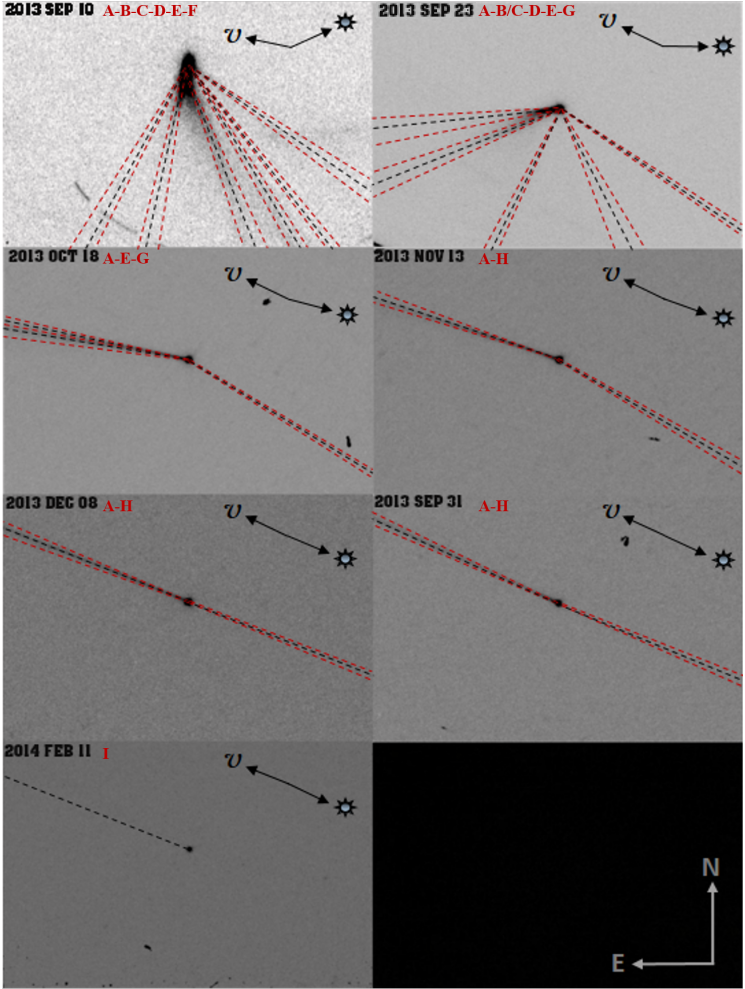}
    \caption{Composite images of 311P/PanSTARRS. These images were taken during the years 2013 and 2014, and the specific date is marked in the upper left corner of each panel. The arrow marked with N denotes the direction of north, and the arrow marked with E denotes the direction of east. The arrows labeled with $v$ and the solar symbol show the directions of 311P's motion and the solar orientation, respectively. Black dashed line denotes the brightest streamer of each tail, which limits the morphology of the tail together with the two adjacent red dashed borderlines. Due to the fading of the dust tail on 11 February 2014, only the streamer with the signal-to-noise ratio greater than 5 is displayed in the bottom-left panel.}
    \label{fig:fig2}
\end{figure*}

\section{Dust Tail Morphology And Finson-Probstein Analysis}
\label{sec:DustTailMorphology}
It is known that 311P manifest unique cometary characteristics, i.e.~compact coma and several extended tails. 
In order to make a morphological analysis of 311P and estimate the quantity of the dust in the tails, a set of raw data (available on the MAST Archive website \url{https://archive.stsci.edu}) is used in this paper.
For each image of the observation date, ray removal is done within the full range of raw image \citep{joye2003new}.
The cleared image is obtained by interpolating regions of the image affected by rays with the average pixel count regions surrounding the ray \citep{bertin1996sextractor}.
We stack observation images at seven epochs separately to increase the S/N ratio of the dust tails, and also apply the median subtraction algorithms to feature the structures of the outer tail \citep{piccardi2004background}.
Meanwhile, we apply the brightness contours fitting to improve the clarity of tails, with the aim of determining the accurate position angle of the tails \citep{bertin1996sextractor}.

We identify the morphology of 311P's tails from the observation data. Firstly, the boundary of each tail in the image whose features have been enhanced is roughly determined by the visual exploration. 
After that, we evaluate the local signal-to-noise ratio to get the detail of general tail structure,
with the local resolution that was set as $4\times4$ pixels.
Combining the tails' contours in each image, we obtain three streamers representing the boundaries and the brightest line of each tail shown in Figure~\ref{fig:fig2}. 
Note that the dust tail looks very faint from the image observed on 11 February 2014, but we still identify a line whose signal-to-noise ratio reaches a value of 5 over a length of $21.4\arcsec$ at the position angle of 66\degr.
Finally, the position angle and the length of tails are obtained by fitting the morphology of tails to contours.
The Measurement of the tails' features is listed in Table~\ref{tab:2}.

\begin{sidewaystable*}
\centering
\setlength\tabcolsep{8pt}
\caption{Measurements of streamers in the tails}
\label{tab:2}
\begin{tabular}{ccccccccccccc} 
\hline 
&\multicolumn{4}{c}{10 September 2013}& \multicolumn{4}{c}{23 September 2013}& \multicolumn{4}{c}{18 October 2013}\\
\textbf{Tail}& \multicolumn{3}{c}{\textbf{\underline{Position angle of streamer (\degr)}}}& \textbf{\underline{Length (\arcsec)}}&\multicolumn{3}{c}{\textbf{\underline{Position angle of streamer}}}& \textbf{\underline{Length}}&\multicolumn{3}{c}{\textbf{\underline{Position angle of streamer}}}&\textbf{\underline{Length}}\\ 
& upper& brightest& lower&brightest& upper& brightest& lower&brightest& upper& brightest& lower&brightest\\ 
\hline 
A&$238\pm1$&$235\pm1$&$233\pm0.5$&21.5&$236\pm1$&$235\pm0.5$&$234\pm2$&22.7&$234\pm0.5$&$233\pm0.1$&$233\pm1$&24.7\\ 
B&$220\pm1$&$218\pm0.5$&$217\pm0.5$&30.1&$205\pm3$&$201\pm4$&$195\pm3$&35.3&/&/&/&/\\
C&$215\pm0.1$&$213\pm0.5$&$210\pm1$&31.5&/&/&/&/&/&/&/&/\\
D&$203\pm0.1$&$202\pm0.5$&$199\pm2$&26.3&$155\pm2$&$153\pm1$&$151\pm0.5$&37.4&/&/&/&/\\
E&$162\pm1$&$161\pm1$&$159\pm0.5$&25.6&$115\pm0.5$&$112\pm0.5$&$108\pm1$&38.9&$85\pm1$&$84\pm1$&$83\pm0.5$&37.2\\
F&$143\pm2$&$141\pm0.1$&$140\pm1$&18.2&/&/&/&/&/&/&/&/\\
G&/&/&/&/&$98\pm1$&$96\pm1$&$95\pm1$&19.7&$83\pm0.5$&$80\pm1$&$78\pm0.1$&39.8\\
H&/&/&/&/&/&/&/&/&/&/&/&/\\
I&/&/&/&/&/&/&/&/&/&/&/&/\\
\hline
\end{tabular}
\par
\centering
\setlength\tabcolsep{5pt}
\label{tab:tab2-1}
\begin{tabular}{cccccccccccccccc} 
\hline 
\multicolumn{4}{c}{13 November 2013}& \multicolumn{4}{c}{8 December 2013}& \multicolumn{4}{c}{31 December 2013}& \multicolumn{4}{c}{11 February 2014}\\
\multicolumn{3}{c}{\textbf{\underline{Position angle of streamer}}}& \textbf{\underline{Length}}&\multicolumn{3}{c}{\textbf{\underline{Position angle of streamer}}}& \textbf{\underline{Length}}&\multicolumn{3}{c}{\textbf{\underline{Position angle of streamer}}}&\textbf{\underline{Length}}&\multicolumn{3}{c}{\textbf{\underline{Position angle of streamer}}}& \textbf{\underline{Length}}\\ 
upper& brightest& lower&brightest& upper& brightest& lower&brightest& upper& brightest& lower&brightest& upper& brightest& lower&brightest\\ 
\hline
$234\pm0.1$&$233\pm1$&$233\pm0.5$&22.7&$247\pm2$&$246\pm3$&$245\pm3$&25.3&$246\pm3$&$245\pm2$&$245\pm1$&30.1&/&/&/&/\\
/&/&/&/&/&/&/&/&/&/&/&/&/&/&/&/\\
/&/&/&/&/&/&/&/&/&/&/&/&/&/&/&/\\
/&/&/&/&/&/&/&/&/&/&/&/&/&/&/&/\\
/&/&/&/&/&/&/&/&/&/&/&/&/&/&/&/\\
/&/&/&/&/&/&/&/&/&/&/&/&/&/&/&/\\
/&/&/&/&/&/&/&/&/&/&/&/&/&/&/&/\\
$73\pm0.1$&$72\pm0.5$&$70\pm0.1$&25.4&$69\pm0.3$&$68\pm0.8$&$67\pm0.2$&34.2&$67\pm0.2$&$65\pm0.3$&$64\pm0.2$&41.4&/&/&/&/\\
/&/&/&/&/&/&/&/&/&/&/&/&/&$66\pm2$&21.4&/\\
\hline 
\end{tabular}
\end{sidewaystable*}
The physical properties of dust can be derived by analyzing the morphology of 311P. 
The epochs of ejections can be derived from the position angles of the tails, the size range of dust particles can be deduced from the length of a tail, and an upper limit of the duration of the activity can be estimated from the measurement of the widths of the tails.
After leaving the nucleus, the particles are mainly subject to solar radiation pressure and solar gravity. The forces acting on particles are parameterized with the dimensionless constant $\beta$
\begin{equation}
    \beta=\frac{\rm Radiation \ Force}{\rm Solar \ Gravity}
	\label{eq:eq1}
\end{equation}
which is further expressed as a parameter related to dust properties \citep{burns1979radiation}
\begin{equation}
    \beta=\frac{C_{\rm pr}Q_{\rm pr}}{\rho_{\rm d}a}
	\label{eq:eq2}
\end{equation}
Here, $Q_{\rm pr}$ is the scattering efficiency for solar radiation, and is assumed as 1. The variable $\rho_{\rm d}$ is the bulk density of dust in g\,cm$^{-3}$, $a$ is the particle radius in cm,
and the constant $C_{\rm pr}=5.76 \times 10^{3} \mathrm{~g} \mathrm{~cm}^{-2}$. 
The variable $\beta$ is inversely proportional to $a$, so for smaller grains the effect of solar radiation pressure on the motion of particles is stronger.

No dust trail was detected within $40\arcsec$ of the nucleus on 19 March 2015, which is different from the active asteroid 354P (P/2010 A2), whose trail dominated by cm-sized particles was still visible more than four years after the mass-loss event. The absence of the dust trail (\citealt{jewitt2018nucleus}; \citealt{jewitt2013large}) may imply that the activity mechanism of 311P is different from that of 354P.

The upper limit of the grain size can be roughly estimated by analyzing the equilibrium between solar radiation pressure and 311P's gravitational force. Particles larger than the upper limit of the grain size will drop back to 311P's surface \citep{hui2017resurrection}. Assuming that 311P is a prolate spheroid, with a long-axis radius of $b$, and two equal short-axes radii of $a$, and the ratio of these axes is $b/a = 1.3$ \citep{jewitt2018nucleus}, according to Equation (6) of \citet{hui2017resurrection}, the minimum value of $\beta$ that corresponds to the largest dust particle is estimated as 
\begin{equation}
\beta_{\min }=\frac{16 \mathrm{\pi} c G \rho_\mathrm{d}^{2} R^{2} r_{e}}{9\left(1+A\right) S_{\sun} f^{3 / 2}},
\label{eq:largesize}
\end{equation}
Here, $c$ is the speed of light, $G$ is the gravitational constant, $r_{e}$ is the radius of the nucleus, $\rho_{\rm d}$ is the bulk density of the dust that is assumed to be $3.3~\mathrm{g}~\mathrm{cm}^{-3}$, $R$ is the heliocentric distance when 311P is active, $A$ is the geometric albedo that is assumed to be $0.29$ \citep{jewitt2015episodic}, $f$ is the ratio of the axes, and $S_{\sun}$ is the solar constant $(1361~\mathrm{W}~\mathrm{m}^{-2})$. We derive that the minimum value of $\beta$ of ejected particles is about 0.0002, which corresponds to a upper limit of grain size of 5 mm approximately. 

We adopt the Finson-Probstein theory to analyze the dust tails \citep{finson1968theory}. The Finson-Probstein theory proposes the concept of the syndyne-synchrone diagram. A syndyne represents loci of particles with the same value of $\beta$, while a synchrone represents loci of particles released at the same time. In this paper, the ejection speed is assumed to be zero. We first verify the validity of the assumption of the zero ejection speed by using Equation (6) of \citet{agarwal2015hubble}
\begin{equation}
    \frac{E}{m}=\frac{E_{\mathrm{n}}}{m_{\mathrm{n}}}+v_{\mathrm{n}} v_{\mathrm{ej}}+\frac{1}{2} v_{\mathrm{ej}}^{2}+\frac{G M_{\odot} \beta}{R}.
	\label{eq:eq3}
\end{equation}
where the left side of the equation is the energy term of a particle, $E$ is the particle's energy and $m$ is the particle mass. The first item of the right side of the equation represents the energy per unit mass of 311P, $E_{\mathrm{n}}$ is 311P's energy and $m_{\mathrm{n}}$ is the mass of 311P. The variable $v_{\mathrm{n}}$ is the speed of the nucleus, $v_{\mathrm{ej}}$ is the particle's relative velocity to 311P, $G$ is the gravitational constant, $M_{\odot}$ is the mass of the Sun, and $R$ is the heliocentric distance of 311P. The assumption of the zero ejection velocity is valid if the term $v_{\mathrm{n}} v_{\mathrm{ej}}$ is negligible compared to the solar radiation pressure term $\frac{G M_{\odot} \beta}{R}$ by using Equation (7) of \citet{agarwal2015hubble}, i.e.
\begin{equation}
v_{\mathrm{ej}} \ll \beta \frac{G M_{\odot}}{R v_{\mathrm{n}}} = 
\beta\,\times\,2.4\,\times\,10^{4} {\rm m} \, {\rm s}^{-1}
	\label{eq:eq4}
\end{equation}
For 311P, $v_{\mathrm{n}}\sim1.9\times10^{4}$ m\,s$^{-1}$, and $R\sim2$ AU. For the minimum $\beta$ of 0.0002 estimated by Equation (\ref{eq:largesize}), we obtain the minimum value of the right side of Equation (\ref{eq:eq4}) as 4.8 m\,s$^{-1}$. Thus, for dust particles with radius less than 5 mm, if $v_{\mathrm{ej}} \ll 4.8 \, \mathrm{m\,s}^{-1}$, the effect of the initial velocity on the evolution of particles is much smaller than that of solar radiation pressure, i.e.~the zero ejection velocity assumption is valid. Later in this Section, we will demonstrate that the majority of dust particles satisfy $v_{\mathrm{ej}} \ll 4.8 \, \mathrm{m\,s}^{-1}$. 

We estimate the point-spread function (PSF) of each image by measuring the Full Width at Half Maximum (FWHM) of the field star trails, and find that PSF of each image is narrower than one-third of the FWHM of each tail of 311P. Thus, the straight-band morphology of 311P's tails from the observation data can be analyzed by using the syndyne-synchrone diagrams (\citealt{finson1968theory}; \citealt{vincent2014comet}). We also estimate the projected width caused by the initial speed, and find that this value is much smaller than the width of each tail at the observation epoch, which indicates that the effect of the initial speed on the projected width of the tail is negligible. Here we take Tail F for instance, which contains the smallest particle with the largest initial speed among all tails (as shown later in Table \ref{tab:tab3} and Equation (\ref{eq:eq5})). Assuming that the component of the initial velocity perpendicular to the orbital plane is equal to the one in the orbital plane, the distance travelled by the smallest particle of Tail F in the direction perpendicular to the orbital plane due to the initial speed estimated by Equation (\ref{eq:eq5}) during the time interval between emission and observation (10 September 2013) is about 58 km, which is much smaller than the FWHM ($\sim$ 480 km) at its maximum length. Therefore, the upper limit of the duration of each activity can be estimated from the width of the tail by using the synchrone analysis.

\begin{table}
\centering
\setlength\tabcolsep{10pt}
	    \caption{Tail-fitting analysis results for 311P}
	    \label{tab:tab3}
	 \begin{tabular}{ccccc} 
		\hline
		 Tail & Starting date & Upper limit  & Maximum size  & Minimum size \\
		      & of emission    & of duration (days)     & of grains (mm)     & of grains (mm)\\
		\hline
		A&23 Mar&8&38.90&1.95\\
		B&19 July&6&3.97&0.16\\
		C&26 July&3&3.13&0.11\\
		D&07 Aug&3&1.57&0.07\\
		E&26 Aug&5&0.35&0.03\\
		F&01 Sep&2&0.11&0.006\\
		G&12 Sep&5&0.16&0.007\\
		H&23 Oct&5&0.6&0.025\\
        \hline 
    \end{tabular}
\end{table}

\begin{figure*}
    \centerline{\includegraphics[width=1\columnwidth]{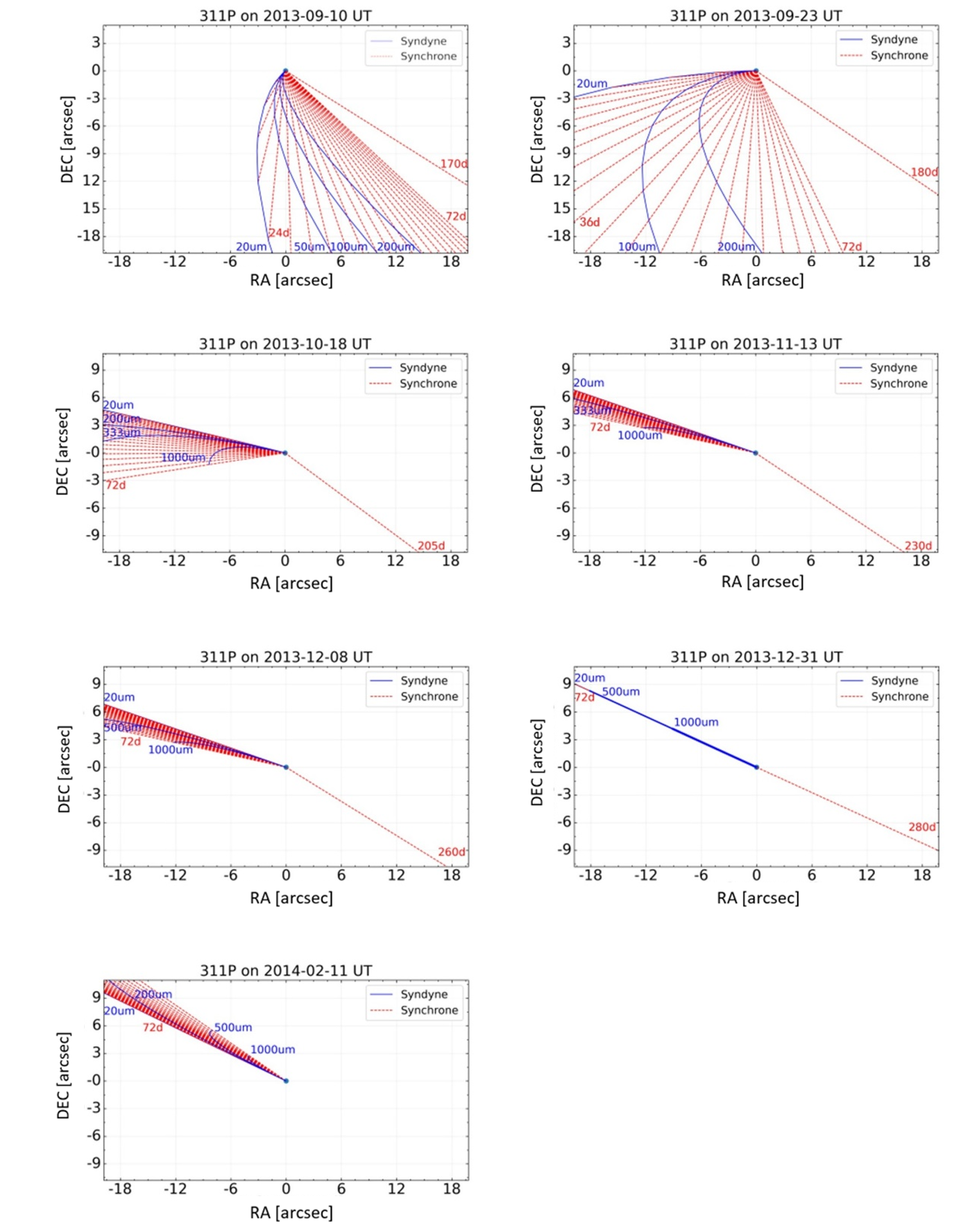}}
    \caption{Synchrones and syndynes of 311P/PanSTARRS. The blue dot at the origin denotes the position of 311P. The synchrones are shown as red dashed lines, with the dust release time prior to the date of the observation labeled as red texts, and the labeled dates increase anti-clockwise with intervals of 3 days. The syndynes are drawn as blue solid lines, with the values of the grain radius (corresponding to different values of $\beta$) labeled as blue texts. Celestial north is up, and east is to the left. The units of RA and DEC are arcseconds (\arcsec).}
    \label{fig:fig3}
\end{figure*}

The position of a dust grain relative to the nucleus depends on both the value of $\beta$ and the difference between the observed epoch and release time (denoted as $\tau$).
To analyse the appearance of the dust tails, we plot a diagram of syndyne (loci of particles with constant $\beta$) and synchrone (loci of particles with constant $\tau$) at various epochs in various dates of observations based on the Finson-Probstein model.
In this diagram, syndynes are calculated for $\beta$ = 0.05, 0.02, 0.01, 0.005, 0.002, 0.001, 0.0005, 0.0002, 0.0001, 0.00005, 0.00002, 0.00001, and each synchrone corresponds the trajectories of particles that were emitted simultaneously before each observation date with a step of 1 days. For better visualization, the more sparsely distributed syndynes and synchrones than what we use are shown in Figure~\ref{fig:fig3}.
To determine starting epochs and ending epochs of dust release and the value of $\beta$ that can describe the dust tails, we fit syndyne-synchrone diagrams to the composite images, and select the syncurves which can best simulate the morphology of tails. The epochs of the dust emission 
and the size range of dust could be inferred by fitting the synchrones
and the syndynes to the boundaries of the tails, respectively. For each tail, we fit all the observations at different epochs with the syndyne-synchrone model. It is found that for the same tail the fitting results by using observations at different epochs are basically consistent with each other.

Inserting values of $\rho_{\rm d} = 3.3\times10^{3}$\,kg\,m$^{-3}$ \citep{jewitt2015episodic} to Equation~(\ref{eq:eq2}), the particle radius $a$ in microns is calculated by the formula by $a=1/\beta$. 
The starting, ending time of the mass-loss events, and the size range of dust particles in each tail derived by the syncurves fitting process are summarized in Table~\ref{tab:tab3}. In the oldest tail A, the size range of dust particles is roughly $2\sim40$ mm, while the radius of the dust particles in the youngest tail H is significantly smaller, ranging from about 0.025 mm to 0.6 mm. Note that the values of starting dates of emission are a bit different from previous results by \citet{jewitt2015episodic}. Possible reasons could be that the value of threshold of signal-to-noise ratio used by us to define the boundaries of dust tails is lower and the boundaries of the tails instead of the brightest line are used to fit the synchrones.

The ejection speed is estimated by the size-dependent velocity fitting formula from \citet{moreno2014intermittent}
\begin{equation}
v_{\mathrm{ej}} = 0.12\beta^{1/8}
	\label{eq:eq5}
\end{equation}
According to Table \ref{tab:tab3}, the smallest grains are 0.006 mm (in Tail F), which gives a value of $\beta_{\mathrm{max}}$ as 0.016. Substituting $\beta_{\mathrm{max}}$ into Equation (\ref{eq:eq5}) yields an upper limit of the ejection speed as $0.09$\,m\,s$^{-1}$, the value of which satisfies Equation (\ref{eq:eq4}), which suggests that zero-velocity assumption is valid. Our estimation of the upper limit of the ejection speed ($0.09$\,m\,s$^{-1}$) is generally consistent with that of \citet{jewitt2015episodic} ($0.03$\,m\,s$^{-1}$).

\section{Tail Surface Brightness}
\label{sec:TailSurfaceBrightness}

We adopt the photometric method by \citet{hsieh2004strange} to analyze the surface brightness profiles of all tails that appear in the images.
For this purpose, composite images of different dates are rotated to align the tails with the horizontal direction, and then the radial profile of each tail is extracted. We use two different sizes of rectangular apertures to measure the brightness of the tails and the sky background. For the tails, the size of each aperture to determine the brightness is set to be 12 pixels along the tail axis and 24 pixels in the direction perpendicular to the tail axis. The sky background is measured with two apertures of size of $12\times12$ pixels located above and below the measured aperture of the tails, and then is subtracted from the tail images. The illustration of this scheme can be found in Figure 7 in \citet{hsieh2004strange}.

For each tail, we measure the tail surface brightness along the tail axis within $18\arcsec$ ($15000\sim30000$ km) away from the nucleus.
The tails' surface brightness profiles are measured in counts per aperture width along the length of the tails, integrated across the perpendicular direction, normalized to the geocentric distance of one astronomical unit, and then further normalized to the ratio of the average count rate from tails to that from 311P's nucleus, which is measured with a $5\times5$ pixel box.
This procedure facilitates the comparison of the brightness profile of the same tail between different observation dates. The value of displayed uncertainty is originated from the root mean square variation measured in the sky background.
The readers are referred to Section 4.1 of \citet{hsieh2004strange} for the details of this photometric method.

The radial brightness profiles of different tails for seven dates of observations are shown in Figure~\ref{fig:fig4}. It is seen that the change in intensity is not exactly proportional to the distance from the nucleus.
We describe the characteristics of eight tails' brightness profiles of Figure~\ref{fig:fig4} as follows:

\begin{sidewaysfigure}
\centering
\includegraphics[width=0.9\linewidth]{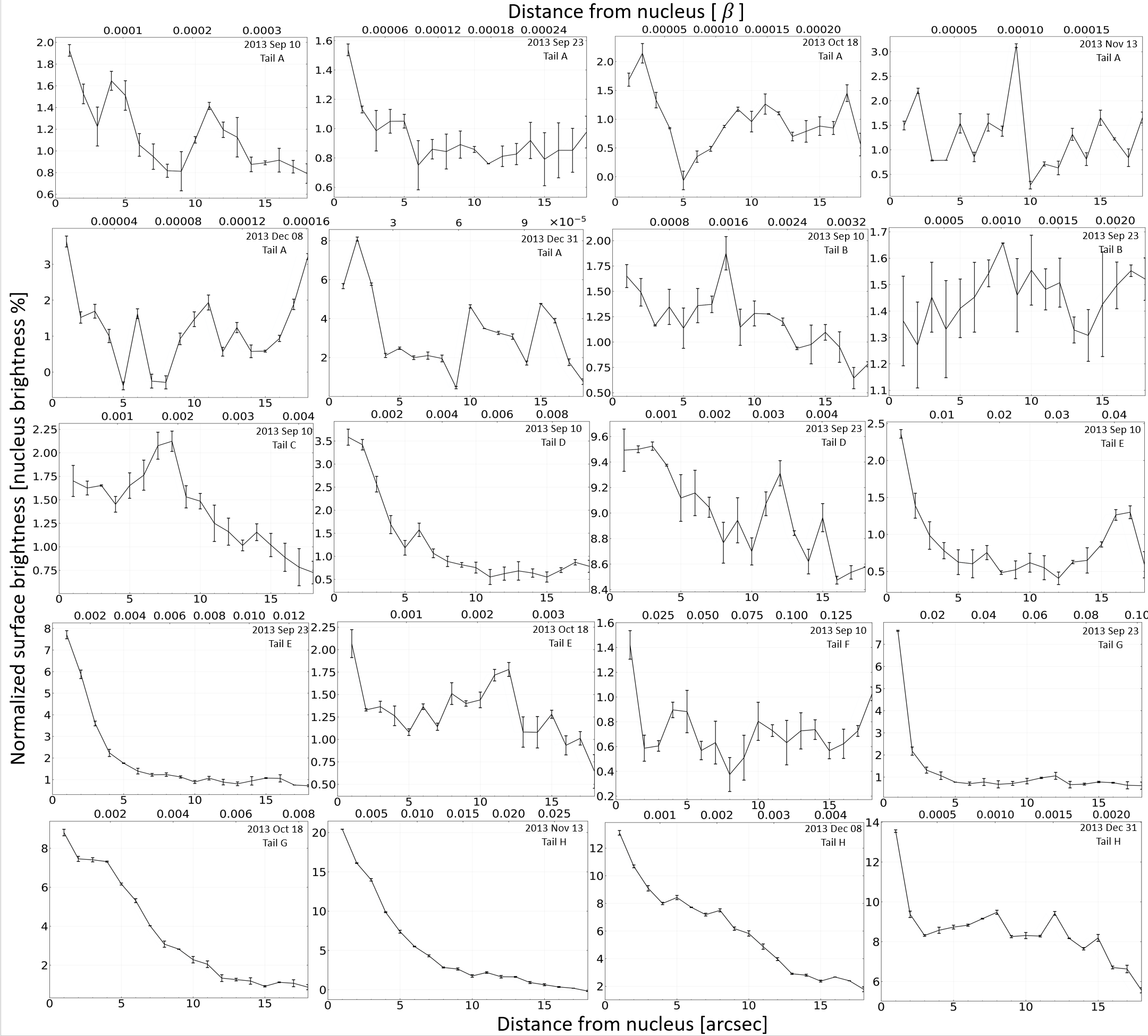}
\caption{Surface brightness profiles of dust tails that are normalized to the brightness of the nucleus from September to December 2013. The dust tail (Tail I) observed on 11 February 2014 is not considered due to its low signal-to-noise ratio. The upper and lower horizontal axes show the radial distance from the nucleus in units of the radiation pressure parameter $\beta$ and arcseconds, respectively.}
\label{fig:fig4}
\end{sidewaysfigure}

Tail A: The brightness distribution $\sum(l)$ of the tail is described by a power law function as $\sum(l)\propto l^{s}$, where $l$ is the angular distance along the tail from the nucleus in arcseconds as seen from the observer.
On 10 and 23 September 2013,
the brightness distribution of Tail A can be characterized by
a shallow power-law function with index $s=-0.25\pm0.05$.
The exponent gradually reduced to 0 in the follow-up observations
although the integrated brightness increased significantly on 31 December 2013.
The peak of the curve at the radial distance of $9\arcsec$ (12000 km) on 8 December is produced by incomplete removal of cosmic rays.

Tail B: Detected on 10 and 23 September, 2013. On 10 September 2013, the index $s=-0.35\pm0.07$. The brightness profile of the dust tail became flatter and
the integrated brightness decreased slightly on 23 September 2013.

Tail C: On 10 September 2013, the brightness profile of Tail C is similar to that of Tail B. On 23 September 2013, Tail C is difficult to distinguish from Tail B.

Tail D: At the first epoch of observation on 10 September 2013, the brightness followed a power law distribution with a slope of about -0.5 within the measurable distance.
On 23 September 2013, the brightness of the dust tail remained at a high level without obvious fluctuations.

Tail E: On 10 September, the surface brightness followed a broken power law distribution, i.e.~the slope of the brightness profile was close to $-1.58\pm0.09$ for the inner tail ($1\arcsec\sim10\arcsec$),
while the slope was flatter ($\sim$-0.4) for the outer tail.
The bulge at $16\arcsec$ (13000 km) attributed to a field star hidden under the tail.
In the latter observation on 23 September,
it was found that the exponent $s$ decreased to $-1.81\pm0.11$ within a distance of $7\arcsec$ (6000 km) from the nucleus.

Tail F: Appeared only in the image observed on 10 September 2013.

Tail G: For Tail G, the slope of the brightness profile decreased with increasing distance from the nucleus, and the integrated brightness increased unusually from 10 to 23 September 2013.
On 13 November 2013, Tail G was almost merged with Tail H.

Tail H: The streamers with above-average signal-to-noise ratios were detected in the direction of the Tail G on 13 November, 8 and 31 December of 2013.
On 13 November, the brightness profile satisfied a power law distribution with the index $s=-0.81\pm0.06$ within a distance of $18\arcsec$ (21000 km) from the nucleus, and the slope reduced to $-0.62\pm0.04$ on 8 December and $-0.21\pm0.02$ on 31 December.
The integrated brightness of the dust tail showed a trend of gradual decrease with the increase of the observation time.

Usually, the brightness profiles of the dust tail at further distance from the nucleus were flatter than that of the coma's at nearer distances from the nucleus \citep{rosenbush2017spatial}.
Similarly, for all observing epochs about 311P, the brightness distribution of the outer tail was flatter than that of the inner tail. It is also found that our results are generally consistent with Figure 10 of \citet{jewitt2015episodic} where the surface brightness profiles of Tails D, E, G and H at certain selected epochs were shown.

\section{Total Mass Of Dust In The Tails}
\label{sec:QuantityOfDust}

In this paper, the total dust mass of each tail is approximated as the dust mass within the region of $18\arcsec$ ($15000\sim30000$ km) from the nucleus, where most dust particles of each tail are located. The dust quantity in the first observation of each tail could give the best approximation of the total dust mass compared to the later observations, because over time some dust particles are pushed away from the region ($18\arcsec$ relative to the nucleus) within which we calculate the dust mass. Besides, the signal-to-noise ratio of the first observation is the highest (the error introduced by the sky background is lowest). Thus, we choose the earliest observation for each tail to analyze. We have found that the value of the size distribution indices calculated by using the later observations are consistent with the one by using the first observation.

\begin{figure*}	
    \plotone{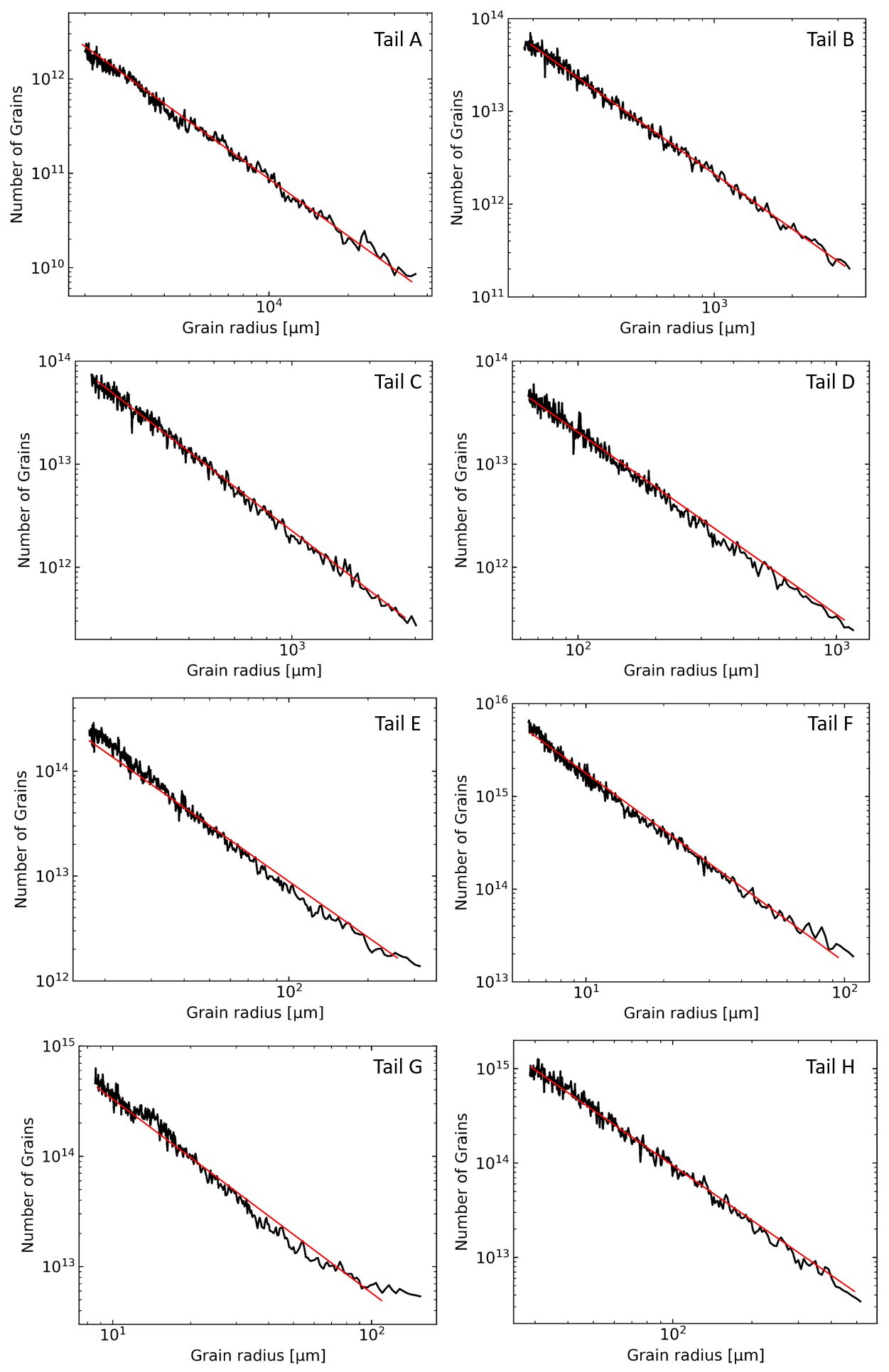}
    \caption{The number of particles vs.~grain radius (black lines). The number of grains is integrated across the whole tail in the direction perpendicular to the tail in the photometry aperture. The red lines are linear fit to the dust size distributions, and the power indices of the distribution functions are, from tails A to H, -2.29, -1.97, -1.88, -1.78, -1.57, -2.11, -1.45 and -1.67, respectively.}
    \label{fig:fig6}
\end{figure*}
In order to estimate the total mass of dust in the tails, we use a sequence of linear apertures perpendicular to the tail with the size of $1\times24$ pixels to evaluate the flux along the tail axis.
By employing the Equation (2) of \citet{jewitt2019initial}, the effective scattering cross-section per pixel, ${C}_e [{\rm km}^2]$, is related to the absolute magnitude of each pixel, by
\begin{equation}
C_{e}=\frac{1.5 \times 10^{6}}{A} 10^{-0.4 H_{\rm pix}}
	\label{eq:quadratic}
\end{equation}
where $H_{\rm pix}$ is the absolute magnitude of each pixel, which is derived from the heliocentric and geocentric distance, the phase angle, the photometric zero-point, and the flux in CCD ADUs per pixel. 

\begin{figure}
\centering
	\includegraphics[width=0.6\columnwidth]{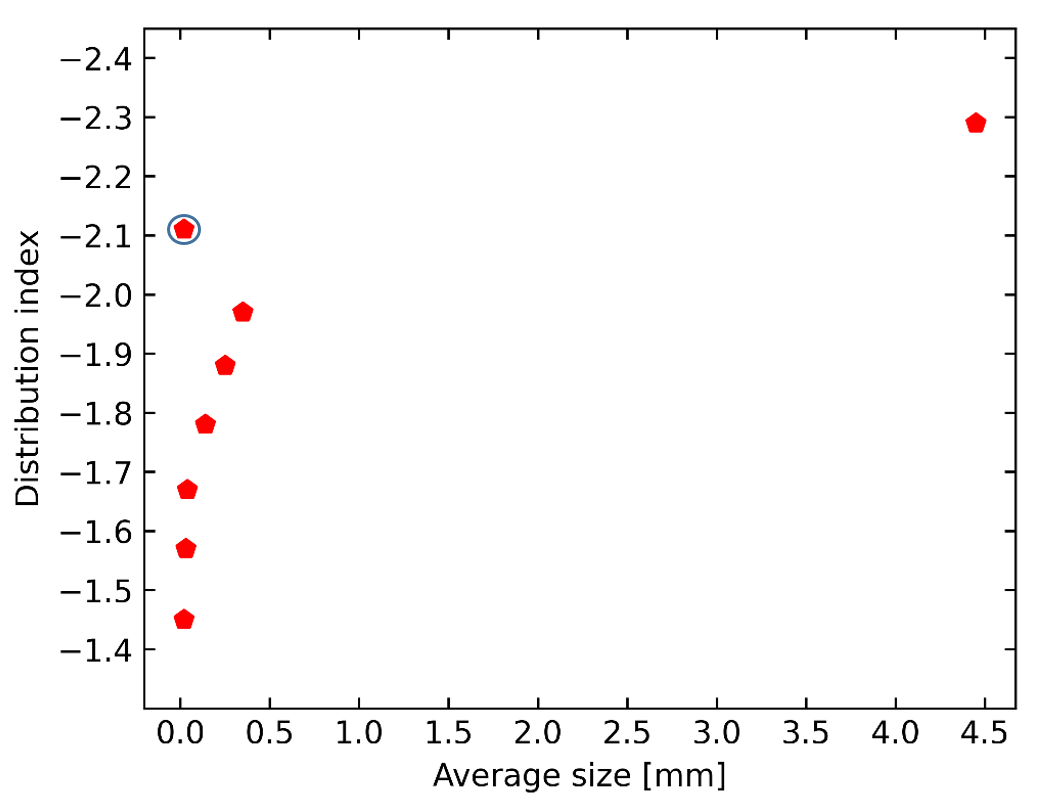}
    \caption{Size distribution index as a function of the average radius of the particles. The eight tails are indicated by red pentagons. The irregularity of the circled point (Tail H) may attribute to the obstruction by the coma.}
    \label{fig:fig9}
\end{figure}

\begin{figure*}
\centerline{\includegraphics[width=\columnwidth]{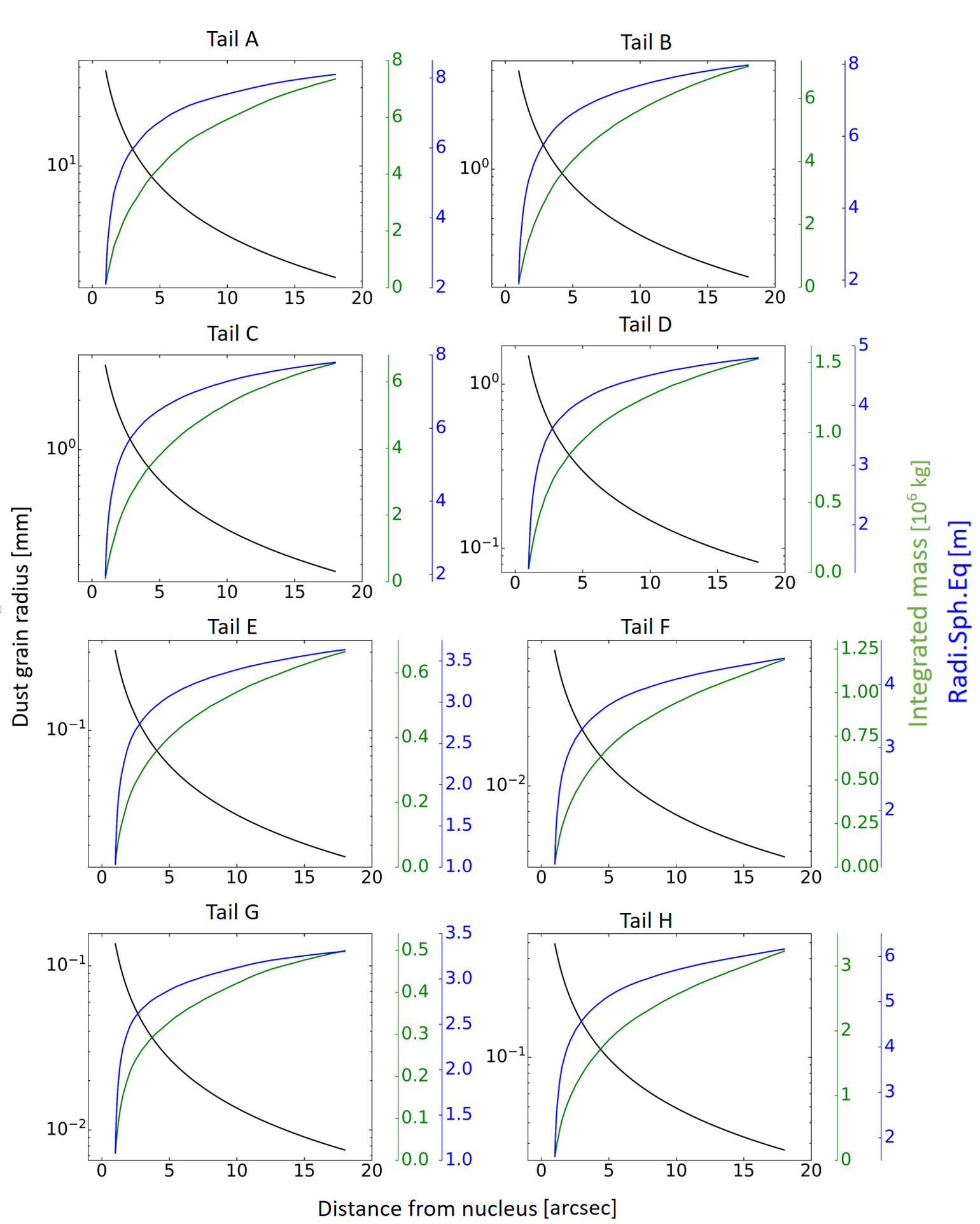}}
    \caption{The average grain size, the integrated mass of tails and the radius of an equal-mass sphere. Each graph contains the following information: (1) The average grain size along the radial distance of the tail from the nucleus (black line). (2) The total mass of dust obtained by integrating the mass of particles along the radial distance of the tail from the nucleus (green line). (3) The radius of an equal-mass sphere, where the mass is obtained by integrating the mass of particles along the radial distance of the tail from the nucleus (blue line).}
    \label{fig:fig5}
\end{figure*}

We adopt the photometric method by \citet{hainaut2012p}, which is summarized as follows. The total cross-sectional area of a single aperture is obtained by adding the contribution of each pixel in this aperture. The tails' brightness distribution as a function of the distance to the nucleus constitutes a “grain radius spectrum” of the dust, and the total cross-sectional area in each single aperture is converted to the number of dust grains at the given distance from the nucleus. Two assumptions \citep{hainaut2012p} are made on the physical properties of dust particles:

1. Based on the Finson-Probstein model used in Section \ref{sec:DustTailMorphology}, tails of 311P are considered as several collections of synchrones. 

2. All particles contained within a single aperture in the dust tails are assumed to be of the same grain size.

\begin{figure}
\centering
	\includegraphics[width=0.6\columnwidth]{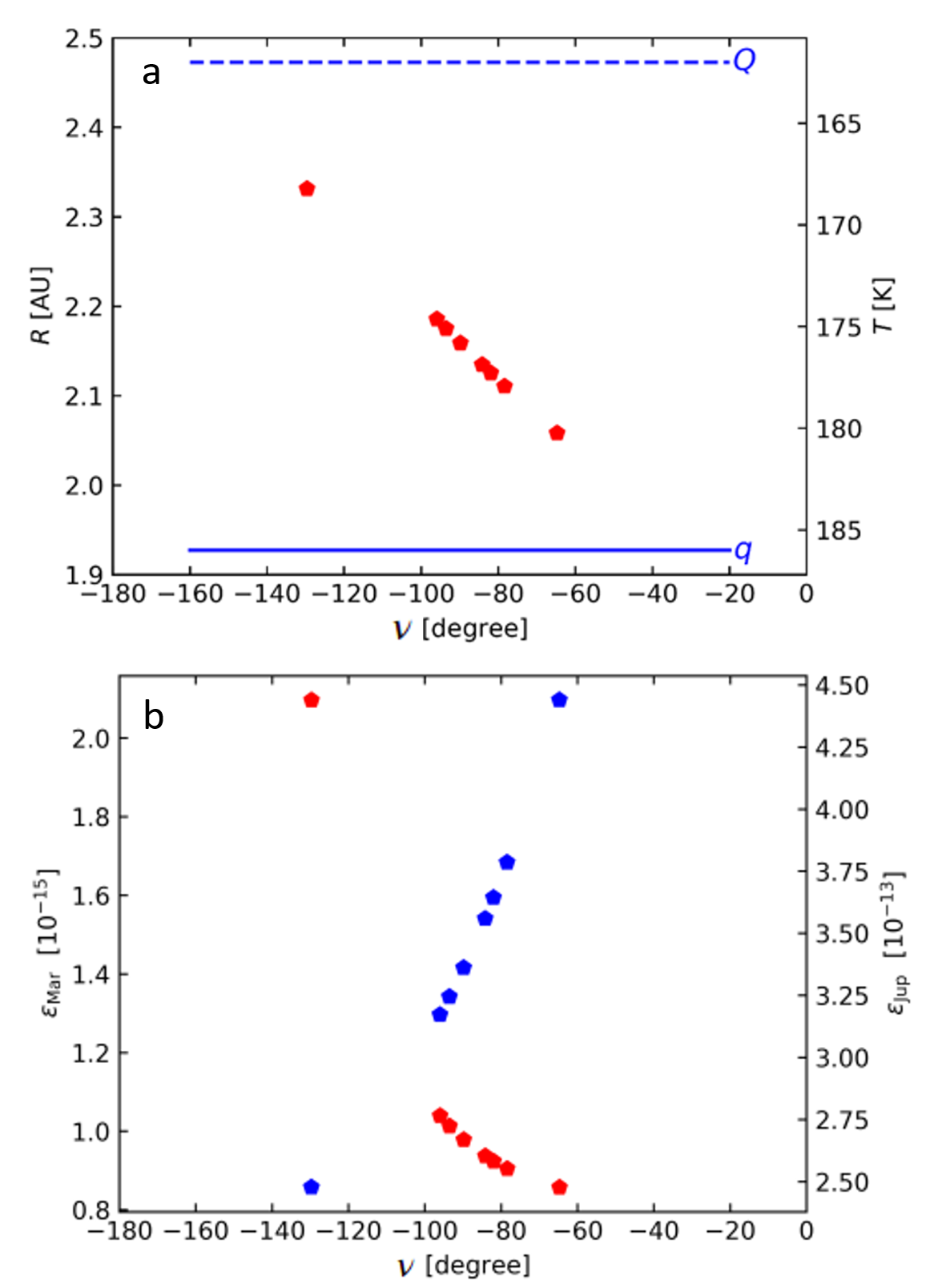}
    \caption{a) The heliocentric distance and the surface temperature as functions of the true anomaly of 311P at the starting time of emission (see Table \ref{tab:tab3}). The emission activities are indicated by red pentagons. Heliocentric distance and surface temperature of 311P when it passed through perihelion $q$ (blue solid line) and aphelion $Q$ (blue dashed line) are shown. b) Dimensionless strengths of the tidal effects caused by Mars (red pentagon) and Jupiter (blue pentagon) at the surface of 311P as functions of the true anomaly of 311P at the starting time of emission. Here, the dimensionless strength of the tidal effect is the ratio of the tidal force caused by the planet relative to the internal gravity of 311P at the surface of 311P (Equation (\ref{eq:varepsilon})).} 
    \label{fig:fig7}
\end{figure}

Following the above method, the number of particles for different grain radii in eight tails are determined and shown in Figure \ref{fig:fig6}. A power-law relation $n(a) \mathrm{d}a\propto a^{q}\,\mathrm{d}a$ is fitted to the distribution of the grain sizes of tails in Figure \ref{fig:fig6}, leading to exponents of $q=-2.29, -1.97, -1.88, -1.78, -1.57, -2.11, -1.45, -1.67$ for the tails from A to H, respectively. We find that for all tails except Tail H, the dust size distributions are steeper for the tails with larger average grain size (Figure~\ref{fig:fig9}). 
Besides, the dust distribution profiles of 311P's tails are similar to those of the two tails of (6478) Gault, of which the dust size distribution indices are $\sim-1.64$ and $\sim-1.70$ in 2018, respectively \citep{kleyna2019sporadic}.

The mass of particles in the tails, which are obtained by integrating dust grains of each single aperture starting from the nucleus, are shown in Figure \ref{fig:fig5}. It is seen that compared to the oldest tail A, the total mass of the youngest tail H is by an order of magnitude smaller. From Figure \ref{fig:fig5} we estimate that 311P ejected at least $3\times10^7$ kg of dust between March and October in 2013, and the total mass of ejecta is equivalent to a sphere with a radius of 13 meters.
Our estimation of the total mass is by an order of magnitude larger than that of \citet{jewitt2015episodic}. The main reason for this difference is that \citet{jewitt2015episodic} estimated the quantities of dust particles within a radius of $5.9\arcsec$ around the nucleus, while we estimate the total mass of dust grains populating all tails within $18\arcsec$ around the nucleus. In other words, our estimation includes small and medium-sized particles at further distances from the nucleus.
It was estimated by \citet{hainaut2014continued} that the total mass of dust for particles smaller than 1.5 mm is about $2.9\times10^7$ kg, which agrees well with our results.

Mass loss rates are estimated by combining with an upper limit of the duration of each eruption in Section \ref{sec:DustTailMorphology}, 
which are $11.7$\,kg\,s$^{-1}$ for Tail A, $13.6$\,kg\,s$^{-1}$ for Tail B, $30.3$\,kg\,s$^{-1}$ for Tail C, $5.8$\,kg\,s$^{-1}$ for Tail D, $1.56$\,kg\,s$^{-1}$ for Tail E, $4.6$\,kg\,s$^{-1}$ for Tail F, $1.08$\,kg\,s$^{-1}$ for Tail G, and $6.6$\,kg\,s$^{-1}$ for tail H, respectively.

\section{Discussion}
\label{sec:Discussion}
In this Section the possible activity mechanisms of 311P are discussed. We first examine the possibility of ice sublimation as the driven mechanism of 311P's activities. 
The surface temperature of a main-belt comet is associated with its orbital location \citep{snodgrass2017main}. Assuming the Sun as a black body, the equilibrium surface temperature of an asteroid is derived as
\begin{equation}
 T=\left[(1-A) \frac{r_{\rm sun}^{2}}{4 R^{2}}\right]^{\frac{1}{4}} T_{\rm sun} 
	\label{eq:temperature}
\end{equation}
where $r_{\text {sun }}=6.957 \times 10^{8} \mathrm{~m}$ is the solar radius, and
$T_{\rm sun}=5777$\,K is the effective temperature of the Sun. From Equation (\ref{eq:temperature}) it is estimated that 311P experiences temperatures ranging from $\sim162$ K at aphelion to $\sim186$ K at perihelion. Consequently, the active asteroid 311P, located within 5 AU from the Sun, is subjected to temperatures that is not low enough for water ice to form \citep{snodgrass2017main, chandler2019six}.
Besides, Figure~\ref{fig:fig7}a shows the variations of the heliocentric distance and the surface temperature as functions of the true anomaly of 311P at the starting time of emission, and it is found that the sequence of the activities began near aphelion and ceased before perihelion, which is different from the sublimation-driven activity. 

We also examine the effects of the tidal force caused by Mars and Jupiter. A rough estimate of the tidal effect is obtained by a dimensionless parameter $\varepsilon$ 
\begin{equation} \label{eq:varepsilon}
\varepsilon = \frac{F_\mathrm{{T}}}{F_\mathrm{{G}}}
\end{equation}
where $F_\mathrm{{T}}$ and $F_\mathrm{{G}}$ are the tidal force of the planet and the internal gravity of 311P at the surface of 311P, respectively. Figure~\ref{fig:fig7}b shows that at the starting time of emission the strength of the tidal effect caused by Mars is only in the order of $10^{-2}$ of that caused by Jupiter, and the effect of Jupiter's tidal force on 311P is only in the order of about $10^{-13}$ of the internal gravity of 311P. Thus, it is unlikely that 311P's activities are correlated with tidal forces caused by Jupiter and Mars.

Another possible driven mechanism is impact. 311P had experienced multiple eruptions, and the likelihood of 311P being hit repeatedly within eight months is low. However, we cannot exclude the possibility that an initial impact-driven event on 311P destabilized the structure of the nucleus and triggered the subsequent activities.

It has been suggested by \citet{jewitt2013extraordinary} and \citet{jewitt2015episodic} that disruption due to rotational instability of the nucleus is a possible origin for the emission activities of 311P. There is no strong evidence against this activation process. Besides, the indices of the dust size distributions of the tails (Figure \ref{fig:fig9}) are close to that of the power-law distributions of the self-organized critical sandpiles \citep{laurson2005power}. We speculate that 311P may be initially activated by rotational instability or impacts in March 2013, and several avalanches of the dust distributed on the surface were triggered and part of the avalanched dust debris was eventually detached from the nucleus.

\section{Conclusions}
\label{sec:Conclusions}
311P experienced several mass-loss events during 2013. To derive the dust environment around the nucleus, we analyze observation images about dust tails of 311P obtained from the Mikulski Archive for Space Telescopes.
The main conclusions of this paper are listed as follows:

1. The position angles of 311P's tails ranged from $64.0\degr$ to $238\degr$. The longest dust tail is Tail H observed on 31 December 2013, with a length of $41.4\arcsec$ ($\sim71000$ km). The shortest dust tail is Tail F observed on 10 September 2013, with a length of $18.2\arcsec$ ($\sim15000$ km).

2. A syndyne-synchrone diagram analysis shows that the estimated upper limits of the durations of the mass-loss events ranged from 2 days (Tail F) to 8 days (Tail A).
The upper limit of the grain radius that dominate tails is 38.9 mm and the corresponding lower limit is 6 $\mathrm{\mu m}$.

3. For all tails, the brightness profiles followed a power law distribution with the index ranged from approximately -1.81 to 0, and the size distribution indices ranged from -2.29 to -1.45. The mass quantities of particles in different tails ranged from 0.5 to 8 $\times10^6$\,kg, and the total mass of dust is $m\sim3\times10^7$\,kg. The average mass-loss rate is approximately $1.59$\,kg\,s$^{-1}$.

4. Analysis of the activities of 311P shows that the possibilities of sublimation, continuous impacts or tidal forces of planets as the origin of the activities can be ruled out. Activation by rotational instability remains a possibility without strong evidence against it. 

\section*{Acknowledgements}
This work was supported by the National Natural Science Foundation of China (No.~12002397 and 12311530055), the National Key R\&D Program of China (No.~2020YFC2201202 and 2020YFC2201101), the grants from The Science and Technology Development Fund, Macau SAR (File No.~0051/2021/A1), and by the Shenzhen Science and Technology Program (Grant No.~ZDSYS20210623091808026). We thank Man-To Hui for helpful discussions and suggestions. We thank Yijun Zou for her productive suggestions. The data used in this work can be found in the Mikulski Archive for Space Telescopes (MAST) at the Space Telescope Science Institute via DOI: 10.17909/59gh-3310.
\section*{Data Availability}

The data used in this work is generated as detailed in the text and will be shared on reasonable request to the corresponding author.

\bibliography{sample631}{}
\bibliographystyle{aasjournal}

\end{document}